\documentclass[11pt]{article}
\usepackage{graphicx}

\def\be{\begin{equation}}
\def\ee{\end{equation}}
\def\ba{\begin{eqnarray}}
\def\ea{\end{eqnarray}}

\begin{document}
\begin{titlepage}
\title{\begin{flushright}\begin{small} LAPTH-Conf-222/14
\end{small} \end{flushright} \vspace{2cm}
An alternative scenario for critical scalar field collapse in $AdS_3$\footnote{Invited talk at ICHEP2014, Valencia July 2014, Parallel Session `Formal Theory Developments'}}
\author
{G\'erard Cl\'ement\thanks{Email:gerard.clement@lapth.cnrs.fr}\\
\small{LAPTh, Universit\'e de Savoie, CNRS, 9 chemin de Bellevue,} \\
\small{BP 110, F-74941 Annecy-le-Vieux cedex, France} \\ Alessandro
Fabbri\thanks{Email:afabbri@ific.uv.es}\\ \small{Centro Studi e
Ricerche Enrico Fermi� Piazza del Viminale 1, 00184 Roma, Italy}
\\ \small{Dipartimento di Fisica dell'Universit\`a di Bologna,
Via Irnerio 46, 40126 Bologna, Italy}\\ \small{Dep. de F\'isica
Te\'orica and IFIC, Universidad de Valencia-CSIC,}\\ \small{C. Dr.
Moliner 50, 46100 Burjassot, Spain}}
\date{}
\maketitle
\begin{abstract}
In the context of gravitational collapse and black hole formation,
we reconsider the problem to describe analytically the critical
collapse of a massless and minimally coupled scalar field in $2+1$
gravity.
\end{abstract}
\end{titlepage}





\section{Introduction}
\label{intro}
The first exact solution of Einstein's field equations, discovered by Schwarzschild in 1916,
\begin{equation}\label{schw}
ds^2=-(1-\frac{2M}{r})dt^2+\frac{dr^2}{(1-\frac{2M}{r})}+r^2d\Omega^2\
\end{equation}
describes an uncharged and non-rotating black hole of mass $M$. As
shown by Birkhoff  in 1923, this solution is the unique static and
spherically symmetric vacuum solution of General Relativity. Its
properties, i.e. the existence of a trapped region ($r<2M$) bounded
by an event horizon ($r=2M$) and with a (central) singularity
($r=0$), are generic properties of black holes even beyond spherical
symmetry.

Black holes form from the gravitational collapse of massive stars.
By the no-hair theorems \cite{isr68, car71, haw72, rob75} time evolution proceeds towards a static
(stationary) black hole solution uniquely characterized by its
conserved charges (mass, charge and angular momentum). At the
threshold of black hole formation, given by the Chandrasekhar limit,
we have a static
(stationary) black hole solution with a finite mass.

\begin{figure}[h]
\centering \includegraphics[angle=0, height=2.4in] {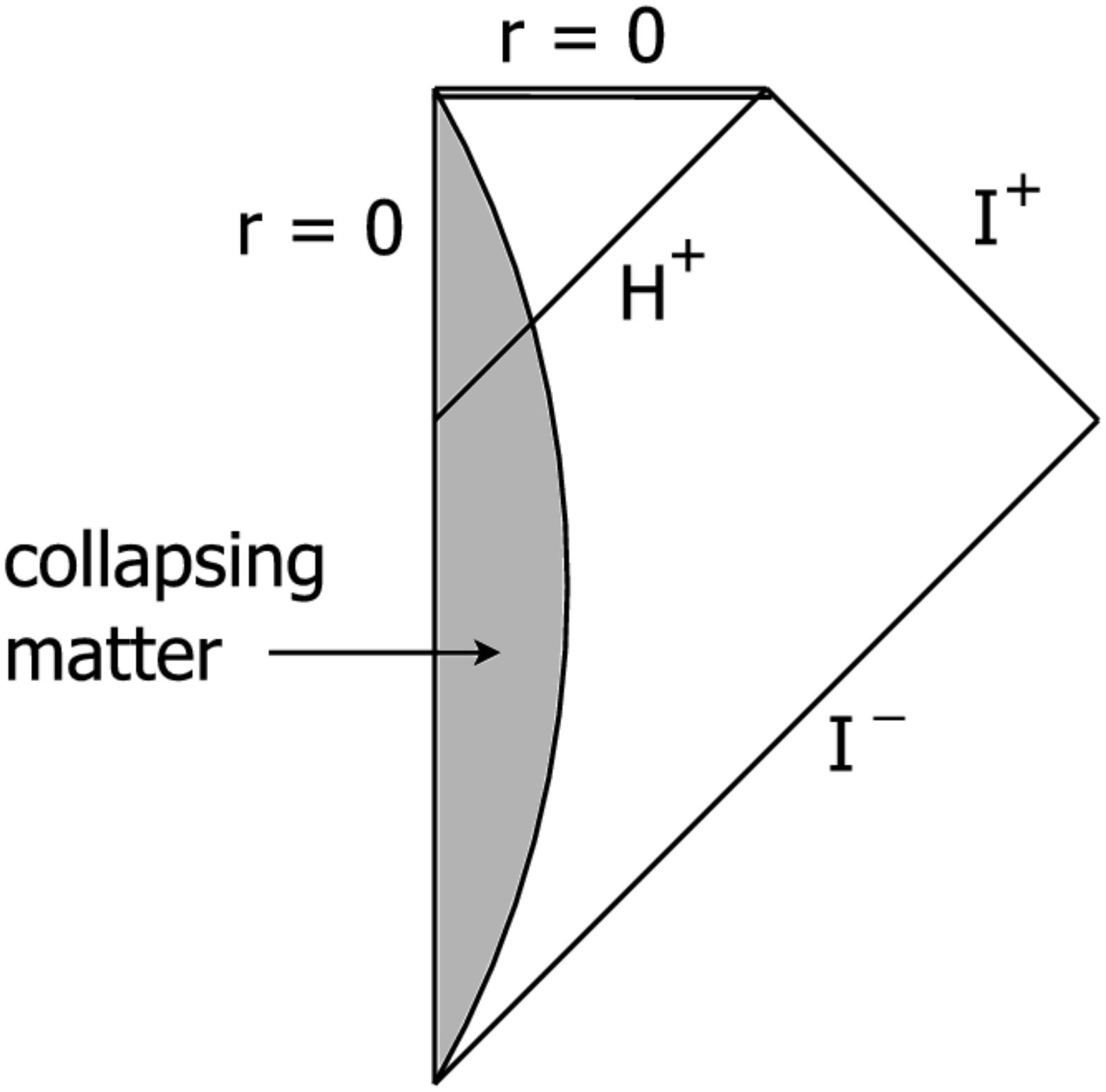}
\caption{Conformal (Carter-Penrose) diagram of gravitational collapse and black hole formation.}
\end{figure}

There is, however, also a critical threshold for black hole formation, discovered by
Choptuik in 1993 \cite{cho93}. He showed that by a fine-tuning of the initial
data one can make arbitrarily small black holes with universal
power-law scaling of the mass and a (continuously or discretely)
scale invariant threshold (critical) solution. There is a striking
similarity between the latter and critical phase transitions in
statistical mechanics \cite{gun02}.

\section{Critical phenomena in gravitational collapse}
\label{crit}

Choptuik \cite{cho93} studied the collapse of a
spherically symmetric massless and minimally coupled scalar field
coupled to 4D Einstein gravity. He considered 1-parameter ($p$)
families of regular initial data and found, empirically, the
critical value $p_*$ discriminating between strong data ($p>p_*$),
in which a black hole forms, and weak data ($p<p_*$), which
disperse. He showed that around the critical point $p=p_*$ there is
a power-law scaling of the black hole mass
\begin{equation}
M\simeq C(p-p_*)^\gamma
\end{equation}
with a universal exponent $\gamma\simeq 0.374$. Moreover, for a
finite time in a finite region of space, near-critical ($p\sim p^*$)
data approach the same universal solution.

Choptuik found that the critical solution is discretely self-similar
(DSS), i.e. it is the same provided we rescale space and time
according to
\begin{equation}
(r,t)=(e^{\Delta}r, e^{\Delta}t) \ ,
\end{equation}
where $\Delta \sim 3.44$ (this phenomena is called scale-echoing). For perfect fluids the
critical solution is continuously self-similar (CSS), i.e. it is
invariant under (infinitesimal and finite) rescaling of space and
time.

Generically, the critical solution has a strong (naked) singularity,
and is characterized by having only one growing perturbation mode.
Analytical approaches consider (CSS or DSS) solutions regular at the
center and at the past light cone of the (naked) singularity (see \cite{gun02} and references therein).

\section{Critical scalar field collapse in 2+1 dimensions}
\label{crit}

In 2+1 dimensions, matter affects space-time only globally and not
locally by producing conical singularities \cite{djh84}. A black
hole solution to the vacuum Einstein equations was found provided we
include a negative cosmological constant ($\Lambda=-\frac{1}{l^2}$)
\cite{btz92}
\begin{equation} \label{btz}
ds^2=-(-M+\frac{r^2}{l^2})dt^2 +
\frac{dr^2}{(-M+\frac{r^2}{l^2})}+r^2d\theta^2\ ,
\end{equation}
which is not asymptotically ($r\to\infty$) Minkowski, as in
(\ref{schw}), but Anti-de Sitter (AdS). The solution (\ref{btz}) is
a black hole (the BTZ black hole) for $M>0$, while for $M<0$
($M\neq-1$) it describes a naked conical singularity and for $M=-1$
it is regular AdS space.

BTZ black hole formation was analysed by \cite{bs00}
in the collision of point-particles and by \cite{ps95} in
the gravitational collapse of a dust ring. No critical solution is
involved in these cases.

Pretorius and Choptuik \cite{pc00} considered the circularly
symmetric collapse of a massless and minimally coupled scalar field
$\phi$ in 2+1 gravity. They considered families of initial data with
length scale $r_0\sim 0.32l$ (so that the effects of the
cosmological constant are suppressed by a factor $0.1$) and tuned to
the threshold of black hole formation on the initial implosion. They
find CSS critical behaviour and power-law scaling (of the maximum
value of the Ricci scalar and of the mass from the apparent horizon)
$(p-p^*)^{2\gamma}$, with $\gamma\sim 1.20\pm 0.05$. Independently,
Husain and Oliver \cite{ho01} found $\gamma\sim 0.81$.

\section{Analytical approach to critical scalar field collapse in $AdS_3$}
\label{ancrit}

Garfinkle \cite{gar01} found a 1-parameter ($n$) family of CSS
solutions to the $\Lambda=0$ equations of motion, regular at the
center, to reproduce the observed critical solution near the
singularity. In appropriate double-null coordinates the metric reads
\begin{equation}
ds^2=-A(v^n+(-u)^n)^{4-2/n}dudv
-\frac{1}{4}(v^{2n}+(-u)^{2n})^2d\theta^2\ . \ \ \
\end{equation}
Regularity of the solutions at the past light-cone of the singularity requires that $n$ is a positive integer.  He found that the solution $n=4$ agrees well with the numerical data.
\begin{figure}[h]
\centering \includegraphics[angle=0, height=1.2in] {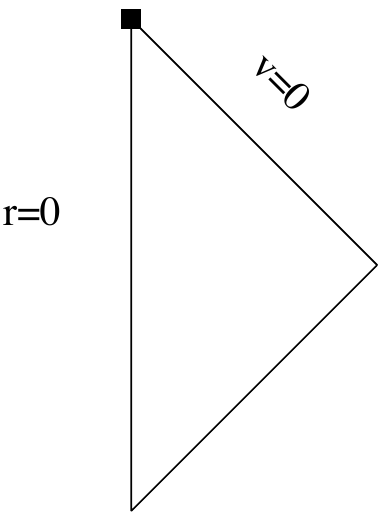}
\caption{Past light-cone of the singularity of the critical solution ($r=g_{\theta\theta}$).}
\end{figure}

Garfinkle and Gundlach \cite{gg02} carried out a linear perturbation
analysis of the Garfinkle solutions. The relevant time parameter
being $\tau=-\ln(-u)^{2n}$ ($\tau=+\infty$ corresponds to the
singularity), the perturbations are expanded in modes
$e^{k\tau}=(-u)^{-2nk}$ which grow when $Re(k)>0$. $k$ is related to
scaling in near-critical collapse: a quantity $Q$ with dimension
(length)$^{s}$ will scale as $|p-p_*|^{s/k}$, i.e.
$\gamma=\frac{1}{k}$.

They imposed regularity condition at the center ($g_{\theta\theta}=0$) and smoothness at $v=0$: they found that the $n=4$ solution has $3$ unstable modes, while $n=2$ has only one unstable mode with $k=\frac{3}{4}$, giving $\gamma=\frac{4}{3}$.  This analysis was extended to $O(\Lambda)$ in \cite{ccf04}.

\section{An alternative scenario}
\label{alt}

A point probably overlooked in the Choptuik and Pretorius analysis
is that the introduction of a point particle (a conical singularity)
left the critical solution unchanged (up to a phase shift in proper
time, related to the mass of the particle). This suggests that the
critical solution, instead of having a regular center, might have no
center at all (as the $M=0$ BTZ vacuum).

Moreover, in the Garfinkle solutions $v=0$ is an apparent horizon:
the critical solution must be something else outside the past
light-cone of the singularity, and it is not clear what are the
correct boundary conditions to be imposed on the perturbations along
this surface.

Recently, Baier, Stricker and Taanila \cite{bst14} (BST) derived,
from a self-similar ansatz, a class of solutions conformal to the 3D
Minkowski cylinder (i.e. with no center). Such solutions belong to a
class of separable solutions \cite{cf02}
\begin{equation}
ds^2=F^2(T)[-dT^2+dR^2+G^2(R)d\theta^2],\ \phi=\phi(T),
\end{equation}
characterized by two parameters $\alpha$ and $b$. When $b$ (the
scalar field strength) vanishes we recover the BTZ solutions with
$\alpha=-M$, while the BST solutions correspond to $b\neq0$,
$\alpha=0$. BST incorrectly suggested that $b=\alpha=0$ leads to the
critical solution: this is not possible since the observed critical
solution has a strong singularity \cite{cf14}.

Unlike the case of the 
Garfinkle solutions we have exact
solutions for $\Lambda\neq 0$ \cite{clefa}. The $\alpha<0$ solutions
are black-hole like, while those for $\alpha>0$ have a center
(regular if $\alpha=1$). Therefore $\alpha=0$ is a candidate
threshold (critical) solution. In the limit $\Lambda=0$ our
solutions take the form
\begin{eqnarray}
ds^2&=&b^2\sinh^2(T)[-dT^2+dR^2+(e^R-\frac{\alpha}{4}e^{-R})]^2d\theta^2\nonumber
\\ \phi&=&\sqrt{2}\ln\tanh(-\frac{T}{2})
\end{eqnarray}
and we see that the center ($\alpha>0$) is sent to $R\to -\infty$
when $\alpha\to0$. They have a singularity at $T=0$, and when
$\alpha=0$ (unlike the Garfinkle solutions) the apparent horizon is
at infinite geodesic distance.

Moreover, the subcritical ($\alpha>0$) solutions near the
singularity are in qualitative agreement with numerical data, and
for a regular center ($\alpha=1$) approximate the $n=1$ Garfinkle
solution.

\section{Open questions}
\label{opq}

The linear perturbation analysis of our solutions indicates that
there is only one unstable growing mode with $k=2$, giving
$\gamma=\frac{1}{2}$ which disagrees with the value $\gamma=O(1)$
from the numerical analysis. Also, our subcritical solution agrees
for $\alpha=1$ with the $n=1$ Garfinkle solution, but the numerical
data are best fir for $n=4$. It will be interesting to see if, along
these lines, one could find a family of solutions which approximate,
near the singularity, the $n=4$ solution, while leading to a
critical exponent $\gamma=O(1)$.




\nocite{*}
\bibliographystyle{elsarticle-num}
\bibliography{martin}

\begin{thebibliography}{00}
\bibitem{isr68}
W. Israel, {\it Commun. Math. Phys.} {\bf 8} (1968), 245
\bibitem{car71}
B. Carter, {\it Phys. Rev. Lett.} {\bf 26} (1971), 331
\bibitem{haw72}
S.W. Hawking,  {\it Commun. Math. Phys.} {\bf 25} (1972), 152
\bibitem{rob75}
D.C. Robinson, {\it Phys. Rev. Lett.} {\bf 34} (1975), 905
\bibitem{cho93}
M.W. Choptuik, {\it Phys. Rev. Lett.} {\bf 70} (1993), 9
 \bibitem{gun02}
C. Gundlach, {\it Phys. Rept.} {\bf 376} (2003), 339
\bibitem{djh84}
S. Deser, R. Jackiw and G. 't Hooft, {\it Annals of Phys.} {\bf 152} (1984), 220
\bibitem{btz92}
M. Ba\~nados, C. Teitelboim and J. Zanelli, {\it Phys. Rev. Lett.} {\bf 69} (1992), 1849
\bibitem{bs00}
D. Birmingham and S. Sen, {\it Phys. Rev. Lett.} {\bf 84} (2000), 1074
\bibitem{ps95}
Y. Peleg and A. Steif, {\it Phys. Rev.} {\bf D51} (1995), 3992
\bibitem{pc00}
F. Pretorius and M.W. Choptuik, {\it Phys. Rev.} {\bf D62} (2000), 124012
\bibitem{ho01}
V. Husain and M. Olivier, {\it Class. Quant. Grav.} {\bf 18} (2001),
L1
\bibitem{gar01}
D. Garfinkle, {\it Phys. Rev.} {\bf D63} (2001), 044007
\bibitem{gg02}
D. Garfinkle and C. Gundlach, {\it Phys. Rev.} {\bf D66} (2002), 044015
\bibitem{ccf04}
M. Cavagli\`a, G. Cl\'ement and A. Fabbri, {\it Phys. Rev.} {\bf D70} (2004), 044010
\bibitem{bst14}
R. Baier, S.A. Stricker and O. Taanila, {\it Class. Quant. Grav.} {\bf 31} (2014), 025007
\bibitem{cf02}
G. Cl\'ement and A. Fabbri, {\it Nucl. Phys.} {\bf B630} (2002), 269
\bibitem{cf14}
G. Cl\'ement and A. Fabbri, {\it Class. Quant. Grav.} {\bf 31} (2014), 098001
\bibitem{clefa}
G. Cl\'ement and A. Fabbri, ``A scenario for critical scalar field
collapse in $AdS_3$'', arXiv:1404.0589 [gr-qc]



\end{thebibliography}



\end{document}